\documentclass[a4paper,amsmath,amssymb,color,nofootinbib,preprint,tightenlines]{revtex4-2}
\usepackage{graphicx}
\usepackage[dvipsnames]{xcolor}
\usepackage[mathscr]{euscript}
\usepackage{slashed}
\usepackage{tablefootnote}
\usepackage[normalem]{ulem}
\usepackage[colorlinks=true,pdfstartview=FitV,
citecolor=blue,linkcolor=purple,urlcolor=Sepia]{hyperref}
\usepackage[capitalise]{cleveref}
\usepackage{orcidlink}
\begin{document}
\baselineskip=17pt \parskip=3pt

\newcommand{\XGH}[1]{{\color{red}XGH: #1}}
\newcommand{\XDM}[1]{{\color{orange}XDM: #1}}
\newcommand{\GV}[1]{{\color{blue}GV: #1}}
\newcommand{\JT}[1]{{\color{purple}JT: #1}}

\title{Light dark-matter window constrained by \boldmath$K^+\to\pi^+$$+$$\not{\!\!E}$}

\author{Xiao-Gang He\,\orcidlink{0000-0001-7059-6311}$,^{1,2,}$\footnote{\email{hexg@sjtu.edu.cn}} Xiao-Dong Ma\,\orcidlink{0000-0001-7207-7793}$,^{3,4,}$\footnote{\email{maxid@scnu.edu.cn}} Jusak Tandean\,\orcidlink{0000-0001-7581-2351}$,^{1,2,}$\footnote{\email{jtandean@yahoo.com}} and German Valencia\,\orcidlink{0000-0001-6600-1290}$^{5,}$\footnote{\email{german.valencia@monash.edu}} 
\vspace{1ex} \\ \it
$^1$State Key Laboratory of Dark Matter Physics, Tsung-Dao Lee Institute \& School of Physics and Astronomy,
Shanghai Jiao Tong University, Shanghai 201210, China \vspace{1ex} \\
$^2$Key Laboratory for Particle Astrophysics and Cosmology (MOE) \& Shanghai Key Laboratory for Particle Physics and Cosmology,
Tsung-Dao Lee Institute \& School of Physics and Astronomy, Shanghai Jiao Tong University, Shanghai 201210, China \vspace{1ex} \\
$^3$State Key Laboratory of Nuclear Physics and
Technology, Institute of Quantum Matter, South China Normal University, Guangzhou 510006, China \vspace{1ex} \\
$^4$Guangdong Basic Research Center of Excellence for
Structure and Fundamental Interactions of Matter, Guangdong Provincial Key Laboratory of Nuclear Science, Guangzhou 510006, China \vspace{1ex} \\
$^5$School of Physics and Astronomy, Monash University,
Wellington Road, Clayton, Victoria 3800, Australia
\vspace{2em} \\
\rm Abstract 
\vspace{8pt} \\
\begin{minipage}{0.99\textwidth} \baselineskip=17pt \parindent=3ex \small
We explore the constraints on new physics from the recent NA62 observation of the kaon decay $K^+\to\pi^+$+$\not{\!\!E}$ with missing energy $\not{\!\!E}$ in the context of a dark-matter (DM) scenario recently used to accommodate the Belle II finding of an enhanced rate of the $b$-meson decay $B^+\to K^+$+$\not{\!\!E}$ compared to the standard-model expectation.
Specifically, assuming that a light real scalar boson $\phi$ plays the role of DM and working in an effective field-theory framework, we study model independently the impact of operators involving $\phi$ and ordinary quarks on the aforementioned transitions over the kaon mode's kinematical mass region of $m_\phi < (m_K - m_\pi)/2 = 177$ MeV.   
Such a DM particle is subject to significant restrictions from the observed relic abundance and from DM direct-detection experiments incorporating the Migdal effect, as well as from indirect searches in cosmic microwave background data and collider experiments, except when its mass is between 110 and 146 MeV. 
We demonstrate that $K^+\to\pi^+\phi\phi$ can saturate the new-physics window in the NA62 result if $m_\phi$ lies in the 110-130 MeV portion of the range left by the DM constraints, thus providing a complementary constraint on this scenario. 
Improved data from future Belle II and NA62 measurements and DM quests can test it more stringently.  In particular, expanding the NA62 signal window into the region that is now removed due to three-body decay background modes could further explore the remaining mass window for this type of invisible particle,  $130 < m_\phi <  177$ MeV.
\end{minipage}}

\maketitle

\newpage

\section{Introduction\label{intro}}

Recently, the NA62 Collaboration  has reported the first observation of kaon decay $K^+\to\pi^+$+$\slashed E$, with missing energy ($\slashed E$) carried away by emitted invisible particles, its branching ratio measured to be \,${\cal B}(K^+\to\pi^+$+$\slashed E)_{\rm exp}=\big(13.0_{-3.0}^{+3.3}\big)\times10^{-11}$ \cite{NA62:2024pjp}, making this mode the rarest one discovered to date. 
In the standard model (SM), neutrinos ($\nu \bar \nu$) act as the invisibles, and the $K^+\to\pi^+\nu\bar\nu$ prediction is remarkably precise \cite{Brod:2010hi} but still suffers from significant parametric uncertainty arising from the input values of the Cabibbo-Kobayashi-Maskawa  angles. This is illustrated by the three different branching-ratio numbers quoted in Ref.\,\cite{NA62:2024pjp}. 
Conservatively, the one with the largest quoted uncertainty, \,${\cal B}(K^+\to \pi^+\nu\bar\nu)_{\rm SM} = (8.4\pm 1.0)\times 10^{-11}$ \cite{Buras:2015qea}, may be adopted for discussion purposes. 
It is less than the measurement by \,$\Delta{\cal B}_K=(4.6^{+3.4}_{-3.2})\times 10^{-11}$. 

This $\Delta{\cal B}_K$ value can be regarded as a potential window, albeit narrow, into new physics beyond the SM. 
Thus, if \,$K^+\to\pi^+\phi\phi$\, could happen with $\phi$ being an invisible new state, this measurement presently offers an additional constraint on it in the mass range \,$m_\phi<(m_K - m_\pi)/2 = 177$~MeV.\, 
As will be shown later, the reported NA62 signal regions are only sensitive to \,$m_\phi<130$~MeV.\, 
Moreover, in the intriguing case that \,$\Delta{\cal B}_K\neq0$\, persists with increasing statistical significance as more data are collected in the future, it may be that $\phi$ is a dark-matter (DM) particle.  
Analogous ideas have lately been proposed~\cite{Felkl:2023ayn,Abdughani:2023dlr,Wang:2023trd,He:2023bnk,Berezhnoy:2023rxx,Datta:2023iln,Altmannshofer:2023hkn,McKeen:2023uzo,Fridell:2023ssf,Ho:2024cwk,Gabrielli:2024wys,Hou:2024vyw,He:2024iju,Bolton:2024egx,Buras:2024ewl,Kumar:2024ivx,Hati:2024ppg,Altmannshofer:2024kxb,Hu:2024mgf,Calibbi:2025rpx,He:2025jfc,Berezhnoy:2025tiw,Ding:2025eqq} to accommodate the recent Belle II measurement~\cite{Belle-II:2023esi} on the $b$-meson decay \,$B^+\to K^+$+$\slashed E$\, with a rate that is higher than the SM expectation.
It is therefore attractive to consider the possibility that the same new physics affects both \,$\Delta{\cal B}_K\neq0$\, and \,$B^+\to K^+$+$\slashed E$,\, with DM mass \,$m_\phi<177$~MeV\, so that the $K^+$ channel with the DM could occur. 
The interactions of the DM candidate must also be able to reproduce the observed relic abundance and comply with bounds inferred from DM direct, indirect, and collider searches. 
Here we carry out a model-independent analysis of this kind of scenario to look for the parameter space at the level of low-energy effective field theory (LEFT) where all these requisites can be fulfilled. 
We arrive at allowed parameter regions, which can be tested further with upcoming data, especially from Belle II, NA62, and the various DM searches, and which were not covered in the earlier $B^+\to K^+$+$\slashed E$\, studies. 
Interestingly, we find, in particular, that the NA62 result partially closes the kinematic window for light DM scalar pairs that is not yet ruled out by relic density and direct-detection considerations.
 
The organization of the paper is as follows. 
In the next section, we discuss the effective interactions of the DM with SM quarks.
In Sec.\,\ref{BKdecays}, we address in greater detail the implications for the $B^+$ and $K^+$ transitions of interest. 
In Sec.\,\ref{dmcons}, we examine the restrictions on the parameter space from various DM searches. 
In Sec.\,\ref{concl}, we give our conclusions.

\section{Effective DM-quark interactions\label{dmint}}

We suppose that the DM particle $\phi$ is a real scalar boson which is a SM-gauge singlet and interacts only in pair with SM quarks due to it being charged under
some symmetry of a dark sector beyond the SM or odd under a $\mathbb{Z}_2$ symmetry which does not influence SM fields. 
This ensures the stability of $\phi$. 
In the LEFT-like framework that we adopt, the lowest-dimension operators describing the $\phi$-quark couplings of interest are given by~\cite{Badin:2010uh,Kamenik:2011vy}
\begin{align}
\mathcal{L}_{qq\phi^2}^{\phi\tt LEFT} & \,\supset\,
\frac{1}{2} \Big[ C_{d\phi}^{S,kl} \big( \overline{d_k} d_l \big) + C_{d\phi}^{P,kl} \big( \overline{d_k} i\gamma_5^{} d_l \big) 
+ C_{u\phi}^{S,uu} (\overline u u)  
+ C_{u\phi}^{P,uu} (\overline u i\gamma_5 u) \Big] \phi^2 \,,     
\label{eq:LEFT}
\end{align}
where summation over $(kl)=(dd,ss,sd,sb)$ is implicit and the $C$s are generally complex, independent constants. 
The hermiticity of this Lagrangian implies that \,$C_{d\phi}^{S(P),kl}=C_{d\phi}^{S(P),lk*}$\, and $C_{u\phi}^{S(P),uu}$ is real. 
Only the $(\bar d_k d_l) \phi^2$ term in Eq.\,(\ref{eq:LEFT}) can supply the $bs\phi^2$ and $sd\phi^2$ couplings that contribute to \,$B^+\to K^+$+$\slashed E$\, and \,$K^+\to\pi^+$+$\slashed E$,\, respectively, as the matrix element of a pseudoscalar quark bilinear between two pseudoscalar mesons 
vanishes.\footnote{If \,$C_{d\phi}^{P,bs}\neq0$\, in Eq.\,(\ref{eq:LEFT}), the $(\bar bi\gamma_5^{}s)\phi^2$ operator could affect the yet-unobserved \,$B\to K^*$+$\slashed E$\, at levels similar to what is required of $(\bar bs)\phi^2$ to explain the $B^+\to K^+$+$\slashed E$ excess without violating their experimental bounds~\cite{ParticleDataGroup:2024cfk}. 
If \,$C_{d\phi}^{P,sd}\neq0$,\, the $(\bar si\gamma_5^{}d)\phi^2$ operator could contribute to \,$K\to\pi\pi$+$\slashed E$\, and \,$K\to\slashed E$~\cite{Kamenik:2011vy}, their SM branching ratios being of order $10^{-13}$ or less~\cite{Geng:1994cw,Littenberg:1995zy,Geng:1996kd,Chiang:2000bg,Marciano:1996wy,Bhattacharya:2018msv}, as well as to analogous hyperon decays~\cite{Li:2019cbk}, all of which still have relatively weak empirical limits~\cite{E787:2000iwe,E391a:2011aa,Geng:2020seh,Su:2020xwt,BESIII:2025kjj}, implying that these kaon and hyperon modes each provide a potentially wide window into new physics~\cite{Li:2019cbk,Geng:2020seh,Goudzovski:2022vbt}.\medskip}
For $\phi$ to be the cosmological DM candidate, we demand that its annihilation generate the observed relic abundance via the freeze-out mechanism and include the $(\bar uu) \phi^2$ and $(\bar dd) \phi^2$ operators. 
Since the $\phi$ mass must be \,$m_\phi<177$\,MeV\, and the latest results of DM direct-detection experiments have translated into strict bounds on sub-GeV DM, especially when the Migdal effect is taken into consideration, it turns out that the impact of the latter can be alleviated by turning on the $(\bar ss)\phi^2$ term as well, which does not affect the relic density. 
These are discussed in more detail below. 
Within an ultraviolet-complete model, such as those of Refs.~\cite{He:2024iju,He:2025jfc}, the different flavor couplings can be related, but we will not make use of such a feature in this study.

\section{\boldmath$B^+\to K^+$$+$$\slashed E$ and $K^+\to\pi^+$$+$$\slashed E$\label{BKdecays}}

In 2023, the Belle II experiment \cite{Belle-II:2023esi} came up with \,${\cal B}(B^+ \to K^+$+$\slashed E)_{\rm exp} = (2.3\pm 0.7)\times 10^{-5}$, which is 2.7$\sigma$ bigger than the SM prediction \,${\cal B}(B^+ \to K^+\nu \bar \nu)_{\rm SM} = (4.43\pm 0.31)\times 10^{-6}$~\cite{Becirevic:2023aov,He:2023bnk}.\footnote{This does not include the tree-level contribution mediated by the $\tau$ lepton~\cite{Belle-II:2023esi}.}
Combining this new measurement with the earlier findings by $BABAR$~\cite{BaBar:2010oqg,BaBar:2013npw}, Belle~\cite{Belle:2013tnz,Belle:2017oht}, and Belle~II~\cite{Belle-II:2021rof} yields the weighted average \,${\cal B}(B^+ \to K^+$+$\slashed E)_{\rm exp} = (1.3 \pm 0.4)\times 10^{-5}$~\cite{Belle-II:2023esi}, which is larger than the SM value by 2.1$\sigma$. 
This could be interpreted as attributable to the presence of the exotic channel \,$B^+ \to K^+\phi\phi$\, induced by the $(\bar bs)\phi^2$ operator with \,$C^{S,bs}_{d\phi}\neq0$\, in Eq.\,(\ref{eq:LEFT}) and DM mass \,$m_\phi<(m_{B^+} -m_{K^+})/2$.\, 
It has been shown~\cite{He:2023bnk,He:2024iju} that this can indeed accommodate the Belle II result if, for example, \,$C^{S,bs}_{d\phi} \in [3,7]/(10^5$\,TeV)\, and $m_\phi\in[0.1,1]$\,GeV. 

If the same underlying new physics is to induce \,$K^+\to\pi^+\phi\phi$\, as well and fill the potential new-physics window $\Delta{\cal B}_K$ in the aforementioned \,$K^+\to \pi^+$+$\slashed E$\, data from NA62, the DM needs to be sufficiently light with mass \,$m_\phi<(m_{K^+}-m_{\pi^+})/2=177$ MeV.\, 
In this range, the \,$B^+\to K^+$+$\slashed E$  result can be accounted for with \,$C_{d\phi}^{S,bs}\sim (2.5-6)/(10^{5}\,\rm TeV)$~\cite{He:2023bnk,He:2024iju}.

For such low-mass DM, we can deal with \,$K\to\pi\phi\phi$\, and the DM annihilation into light hadrons brought about by the LEFT operators in Eq.\,(\ref{eq:LEFT}) by means of chiral perturbation theory. 
This leads to the $\phi$ interactions with light mesons given by
\begin{align} \begin{array}[b]{rl} 
\mathcal{L}_{\phi P} \! & \displaystyle \supset\, \frac{B_0}{2} \phi^2 \Bigg\{
   \Big(C_{u\phi}^{S,uu} + C_{d\phi}^{S,dd}\Big) \bigg( \pi^+ \pi^- + \frac12 \pi^0 \pi^0 \bigg)
  + \Big(C_{u\phi}^{S,uu}+ C_{d\phi}^{S,ss}\Big) K^+ K^-
\\ & \displaystyle ~~~ +~ \Big( C_{d\phi}^{S,dd}+ C_{d\phi}^{S,ss} \Big) K^0 \bar K^0 + \frac{1}{6} \Big( C_{u\phi}^{S,uu}+ C_{d\phi}^{S,dd} +  4 C_{d\phi}^{S,ss} \Big) \eta^2 
+ \frac{1}{\sqrt{3}} \Big( C_{u\phi}^{S,uu} -  C_{d\phi}^{S,dd} \Big) \pi^0 \eta 
\\ & \displaystyle ~~~ +~ \Bigg[ 
C_{d\phi}^{S,sd} \left( \pi^- K^+ - \frac{1}{\sqrt{2}} \pi^0 K^0 - \frac{1}{\sqrt{6}} \eta K^0 \right) + {\rm H.c.}  \Bigg] \Bigg\} \,, \end{array} \label{chiral}
\end{align}
where \,$B_0 = m_{\pi^+}^2/(m_u+m_d)=2.8$ GeV\, with $m_{u,d}$ values from Ref.\,\cite{ParticleDataGroup:2024cfk}.
From the $C_{d\phi}^{S,sd}$ terms, we derive the differential rates of \,$K\to \pi\phi \phi$ \cite{He:2022ljo}\, 
\begin{align} \label{G'K+}
\frac{d\Gamma_{K^+\to \pi^+\phi\phi} }{dq^2} 
& = \frac{\lambda^{1/2} \big(m_{K^+}^2, m_{\pi^+}^2,q^2\big)\, \lambda^{1/2}\big(m_\phi^2,m_\phi^2,q^2\big)\, B_0^2}{512 \pi^3\, m_{K^+}^3\, q^2} \Big|C_{d\phi}^{S,sd}\Big|^2 \,, 
\\ \label{G'KL}
\frac{d\Gamma_{K_L\to\pi^0\phi\phi}}{dq^2} 
& = \frac{\lambda^{1/2} \big(m_{K^0}^2, m_{\pi^0}^2,q^2\big)\, \lambda^{1/2}\big(m_\phi^2,m_\phi^2,q^2\big)\, B_0^2}{512 \pi^3\, m_{K^0}^3\, q^2} \Big({\rm Re}\,C_{d\phi}^{S,sd}\Big)^2 \,, ~~~
\end{align}
where $q^2$ is the squared invariant mass of the $\phi\phi$ pair and \,$\lambda(x,y,z) = (x-y-z)^2 - 4yz$.\,

\begin{figure}[t]  \centering
\includegraphics[width=0.48\linewidth]{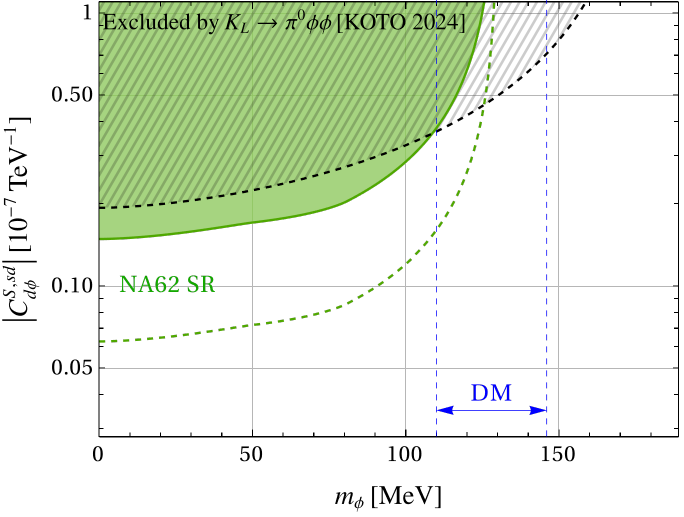}  ~~
\includegraphics[width=0.48\linewidth]{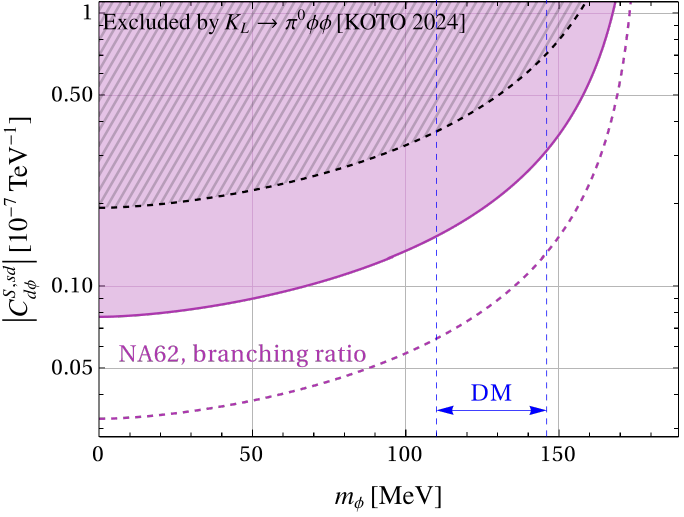}\vspace{-1ex}
\caption{Left: the green-shaded region represents the $\big|C_{d\phi}^{S,sd}\big|$ versus $m_\phi$ parameter space excluded by the latest NA62~\cite{NA62:2024pjp} measurement of \,$K^+\to\pi^+$+$\slashed E$.\, 
The unshaded region between the solid and dashed green curves is where the new-physics window $\Delta{\cal B}_K$  can be populated by  $K^+\to\pi^+\phi\phi$. 
Only the \,$m_\phi\in [110,146]$\,MeV\, range, as indicated, is permitted by the DM relic density requirement. 
The hatched region is excluded by the recent KOTO~\cite{KOTO:2024zbl} search for \,$K_L\to\pi^0$+$\slashed E$\, if $C_{d\phi}^{S,sd}$ is purely real.
Right: the same as the left panel but using only the branching-ratio value reported by NA62~\cite{NA62:2024pjp} and without regard to its signal regions.}
\label{fig:kaon} \vspace{-1ex}
\end{figure}

In the left panel of Fig.\,\ref{fig:kaon}, we have displayed the $\big|C_{d\phi}^{S,sd}\big|$ values versus $m_\phi$ (the unshaded area under the green solid curve) allowed by the NA62~\cite{NA62:2024pjp} measurement on \,$K^+\to \pi^+$+$\slashed E$\, in two signal regions (SRs), the first one (SR1) being specified by \,
$q^2=[0.000,0.010]\,\rm GeV^2$\, and \,$|\textbf{\textit p}_\pi|=[15,35]\rm\,GeV$  and the second (SR2) by \,$q^2=[0.026,0.068]\,\rm GeV^2$\, and \,$|\textbf{\textit p}_\pi|=[15,45]\rm\,GeV$.\,
The right panel of Fig.\,\ref{fig:kaon} is the same as the left, except that only the total branching-ratio result from NA62 has been used. 
In the left (right) panel, the unshaded area between the solid and dashed green (purple) curves corresponds to the aforementioned possible new-physics window $\Delta{\cal B}_K$ in the NA62 data.
The plots in Fig.\,\ref{fig:kaon} also indicate the only mass interval, \,$m_\phi=[110,146]$\,MeV,\, not yet ruled out by the DM direct and indirect searches, to be discussed in Sec.\,\ref{dmcons}.
Additionally, we have drawn on these plots the hatched areas which, if $C_{d\phi}^{S,sd}$ is purely real, are excluded by the limit from the KOTO search for the isospin-related neutral-kaon mode: \,${\cal B}(K_L\to \pi^0$+$\slashed E)< 2.2\times 10^{-9}$ at 90\% confidence level~\cite{KOTO:2024zbl}.
If \,Re$\,C_{d\phi}^{S,sd}=0$\, instead, the KOTO constraint would be absent.

Clearly, the fact that NA62 has to exclude some kinematic regions from their analysis affects significantly the region allowed for new physics that its latest data can accommodate. 
In particular, when combined with the DM constraints, it leaves a modest window for our scenario, namely, that $m_\phi$ has to be between 110 and 130 MeV, as can be viewed in the left panel of~Fig.\,\ref{fig:kaon}.
Nevertheless, on the upside, this means that the $\phi$ interactions in the 130-146~MeV interval remain unrestrained.

\begin{figure}[t] 
\centering
\includegraphics[width=0.48
\linewidth]{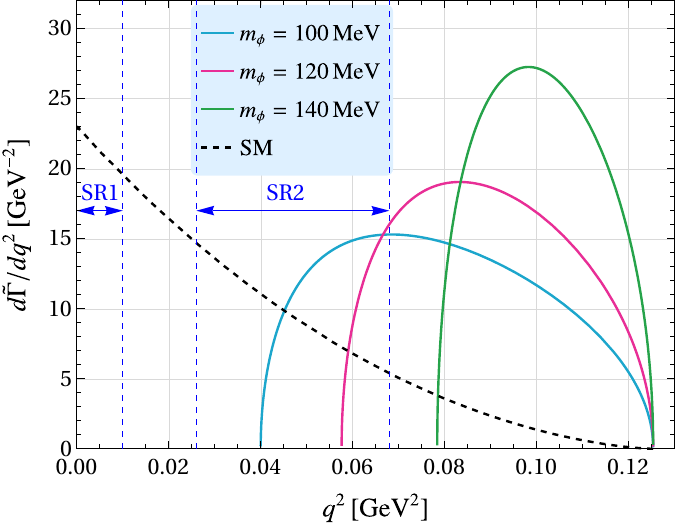} ~ ~ 
\includegraphics[width=0.48
\linewidth]{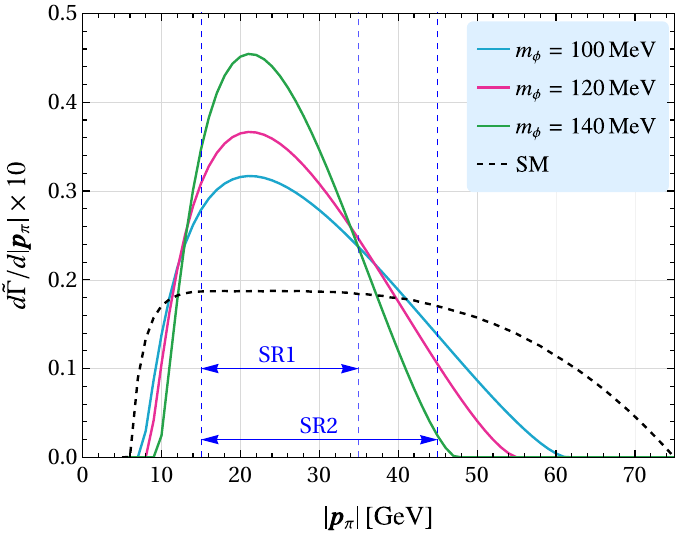}\vspace{-1ex} 
\caption{The normalized distribution of the rate $\Gamma$ of \,$K^+\to \pi^+$+$\slashed E$\, against $q^2$ (left) and the $\pi^+$ momentum $|\textbf{\textit p}_\pi|$ (right) for several benchmark masses of the DM particle.
The labels of the vertical axes are \,$d\tilde \Gamma/dq^2 \equiv (1/\Gamma)d\Gamma/dq^2$\, and \,$d\tilde \Gamma/d|\textbf{\textit p}_\pi| \equiv (1/\Gamma)d\Gamma/d|\textbf{\textit p}_\pi|$, respectively.}
\label{fig:kaondis}
\end{figure}

To understand why the NA62 measurement produces the exclusion region in the left panel of Fig.\,\ref{fig:kaon} instead of the one in the right panel, we have depicted in Fig.\,\ref{fig:kaondis} the kinematic distributions with respect to the squared missing mass $q^2$ and charged-pion momentum $|\textbf{\textit p}_\pi|$ in the lab for the SM and for three different values of DM mass $m_\phi$, along with vertical dashed lines indicating the experimental cuts imposed by NA62. 
The left plot in Fig.\,\ref{fig:kaondis} reveals that the peaking of the new-physics distributions puts them outside the signal region for mass values \,$m_\phi\gtrsim 130$~MeV,\, in which case, as remarked above, they would not be constrained by this measurement and cannot contribute to any observed enhancement over the SM rate. 
It is worth commenting that the exclusion of the large-$q^2$ region by NA62 is intended to remove background from three-body $K^+$ decay modes and is a good compromise when searching for the SM signal, as can be seen in the left panel of  Fig.\,\ref{fig:kaondis}, but should be reconsidered for a study of light, but not massless, new particles. 
Similarly, our scenario does not contribute to this mode in SR1 for the DM masses of interest.
The shape of these two distributions could be used in the future to distinguish between the SM and dark scalars. However, it turns out that, within the relevant SR2, the possible dark scalar models have a kinematic $|\textbf{\textit p}_\pi|$ distribution remarkably similar to its SM counterpart.

\begin{figure}[!t] \centering
\includegraphics[width=0.48\linewidth]{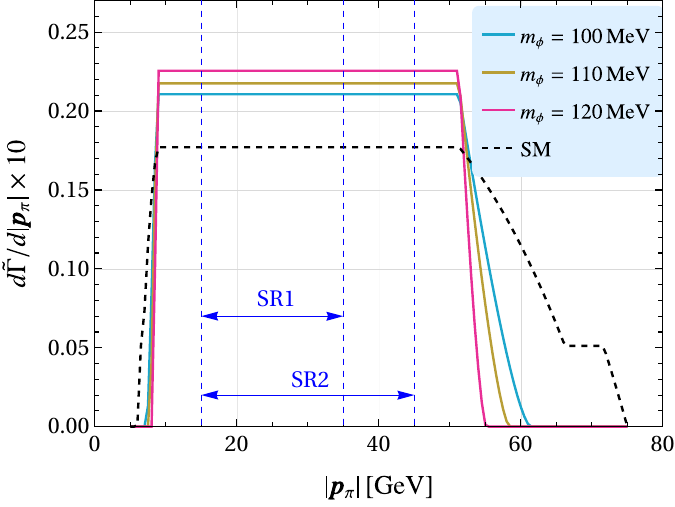}
\vspace{-1ex}
\caption{The normalized distribution against the $\pi^+$ momentum $|\textbf{\textit p}_\pi|$ for several benchmark masses of the DM particle after taking into account the NA62  cuts on $q^2$.}
\label{fig:kaondisq2cut}
\end{figure}

The pion-momentum distribution seen in Fig.\,\ref{fig:kaondis} changes significantly when the $q^2$ limits of the signal region are applied, as displayed in Fig.\,\ref{fig:kaondisq2cut}. This result is surprising at first but can be easily understood because the matrix element squared for the new physics is a constant. This implies that the Dalitz plot distribution for the final state with the pion and a pair of dark scalars simply follows phase space, and all of the features seen in \cref{fig:kaondis} simply reflect the kinematic limits. The NA62 signal region SR2 (which is the relevant one for us) is away from the kinematic limits leading to the result seen in \cref{fig:kaondisq2cut}. The immediate consequence is that the experimental efficiencies will have minimal impact on our signal as it closely resembles the SM within SR2.

\section{Constraints on dark-matter sector\label{dmcons}} 

\subsection{Relic density requirement}

\begin{figure}[h] \bigskip \centering
\includegraphics[width=0.49\linewidth]{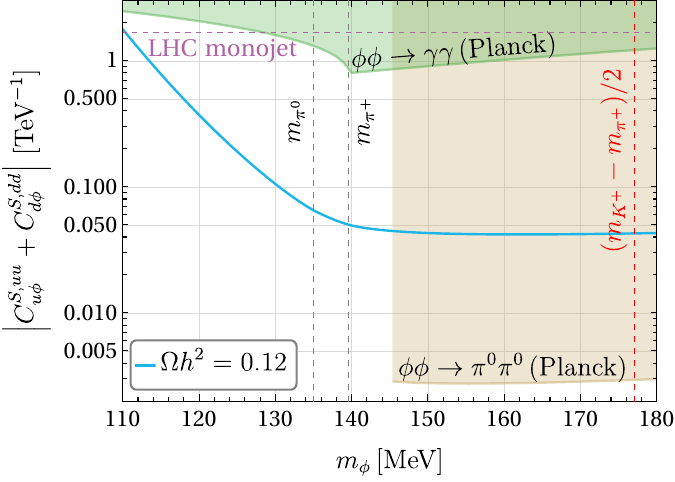} ~ 
\includegraphics[width=0.47\linewidth]{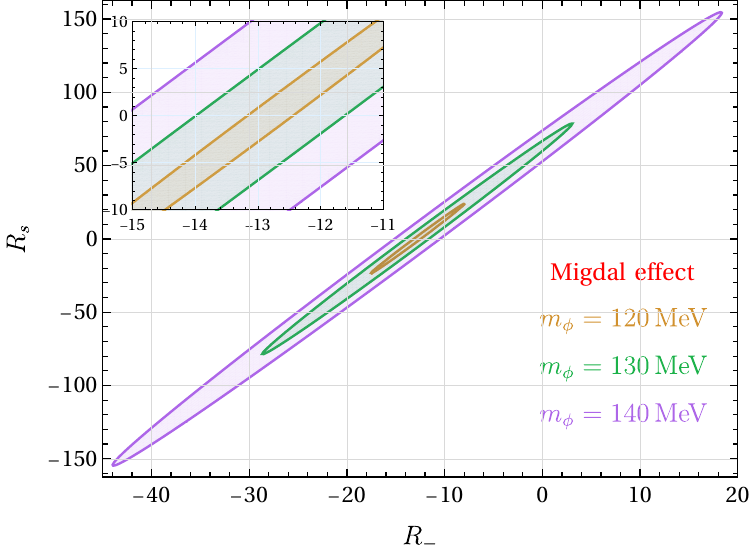}
\vspace{-1ex}
\caption{Left: the cyan solid curve represents the values of $\big|C_{u\phi}^{S,uu}+C_{d\phi}^{S,dd}\big|$ versus DM mass $m_\phi$ that yield the observed DM relic density. 
The brown and green regions are excluded by the CMB constraints~\cite{ODonnell:2024aaw}.  
Right: the $R_-$-$R_s$ regions allowed by the DM direct search result of PandaX-4T \cite{PandaX:2023xgl} incorporating the Migdal effect.
} 
\label{fig:dm}
\end{figure}

We regard $\phi$ as a weakly interacting massive particle,  which is a thermal relic of the early
Universe and undergoes the so-called freeze-out mechanism for the DM production. 
This implies that the observed cosmological abundance of DM can be reproduced from the total self-annihilation cross section of~$\phi$. 
For \,$m_\phi\le177$\,MeV,\, 
the pertinent interactions of $\phi$ are those with pions, already written down in Eq.\,(\ref{chiral}), leading to the cross sections of \,$\phi\phi\to\pi^+ \pi^-/\pi^0\pi^0$, 
\begin{align}
\sigma\big(\phi\phi\to\pi^+ \pi^-/\pi^0\pi^0\big) 
& \,=\, \frac{1}{\mathscr S}
\frac{B_0^2\, \big|C_{u\phi}^{S,uu}+C_{d\phi}^{S,dd}\big|^2}{16\pi\hat s} \sqrt{\frac{\hat s-4 m_{\pi}^2}{\hat s-4m_\phi^2}} \,,
\end{align}
where \,$\mathscr S=1$ and 2\, for the $\pi^+\pi^-$ and $\pi^0\pi^0$ channels, respectively, and $\hat s$ is the squared invariant mass of the $\phi\phi$ pair.
The thermal averages of the corresponding annihilation rates are~\cite{Gondolo:1990dk}
\begin{align}
\langle\sigma v\rangle \big(\phi\phi\to\pi^+\pi^-/\pi^0\pi^0\big) 
& \,=\, \frac{1}{\mathscr S}
\frac{B_0^2\, \big|C_{u\phi}^{S,uu}+C_{d\phi}^{S,dd}\big|^2\, \tilde\eta(x,z_{\pi})}{64\pi m_\phi^2} \,, 
\end{align}
where \,$x\equiv m_\phi/T$\, involves the thermal-bath temperature $T$ of the early Universe, \,$z_{\pi}\equiv m_{\pi}^2/m_\phi^2$,\, and the thermal average integration is captured by the function 
\begin{align}
\tilde\eta(x, z) & \,\equiv\, \frac{4 x}{K_2^2(x)} \int_{\epsilon_{\rm th}}^\infty d \epsilon \, \frac{\sqrt{\epsilon}\, \sqrt{1 + \epsilon - z} }{ \sqrt{1 +\epsilon} }  K_1\big(2 x\sqrt{1+\epsilon}\big) \,,
\end{align}
with $K_i$ standing for the modified Bessel function of order~$i$ and \,$\epsilon_{\rm th}\equiv {\rm max}(0,z_\pi-1)$\, extending the $m_\phi$ region to below the pion mass to include threshold effects~\cite{Gondolo:1990dk}.
Following Ref.\,\cite{He:2024iju}, we determine the $\phi$-quark couplings that can successfully lead to the observed DM relic abundance by matching the total $\langle \sigma v\rangle$ with the required value given in Fig.\,5 of Ref.\,\cite{Steigman:2012nb}. 
In the left panel of~Fig.\,\ref{fig:dm}, we have graphed the cyan solid curve to indicate the resulting $\big|C_{u\phi}^{S,uu}+C_{d\phi}^{S,dd}\big|$ range over the $m_\phi$ region of interest. 

\subsection{Indirect searches}

Since for \,$m_\phi<177$\,MeV\, the DM annihilates mostly into pion pairs, 
these processes can also take place within the DM halo of the Milky Way Galaxy.
Subsequently, the neutral pions emit photons and the charged ones produce them radiatively or via the inverse Compton scattering of their secondary electrons/positrons off the background photons. 
As a result, the DM couplings are subject to constraints from astrophysical x- and $\gamma$-ray observations (from telescopes such as
 INTEGRAL, XMM-Newton, Fermi-LAT, etc.). 
Moreover, when DM annihilation occurs during the epoch of cosmic microwave background (CMB) formation in the early Universe, the energy injected into the cosmic fluid from the annihilation products can alter the CMB anisotropy spectrum. 
Therefore, measurements of CMB temperature and polarization anisotropies also imply restrictions on the annihilation processes. 
Constraints on the annihilation channels into pions and photons from both types of observables have been extensively studied in the literature. 
Since the CMB limits have smaller uncertainties compared to those from the astrophysical x-/$\gamma$-ray observations, and are also the most stringent, we use  
the recent Planck CMB limits on $\langle\sigma v\rangle$ presented in Ref.\,\cite{ODonnell:2024aaw} to probe our parameter space.  

Since the DM is nonrelativistic at the time of CMB formation, the DM velocity effect in $\langle\sigma v\rangle$ is negligible due to the $\tt S$-wave nature of the $\phi\phi$ scattering. 
For \,$m_\pi<m_\phi<177$\,MeV,\, the main annihilation channels are those into $\pi^+ \pi^-$ and $\pi^0\pi^0$,  
\begin{align}
\langle\sigma v\rangle\big(\phi\phi\to \pi^+ \pi^-/\pi^0\pi^0\big)_{\tt I.D.} 
& \,\approx\, \frac{1}{\mathscr S} \frac{B_0^2\, \big|C_{u\phi}^{S,uu}+C_{d\phi}^{S,dd}\big|^2}{32\pi m_\phi^2} \sqrt{1-{m_{\pi}^2\over m_\phi^2}} \,,
\end{align}
where the subscript ``{\tt I.D.}'' indicates that the quantity has been derived for DM indirect searches. 
For \,$m_\phi\le m_\pi$,\, the leading annihilation channel is into two photons, \,$\phi\phi\to \gamma\gamma$,\, due to a charged-pion loop.\footnote{The corresponding charged-kaon loop has a significantly smaller effect.
At tree level, $\phi\phi\to\pi^0\gamma\gamma$\, can occur, mediated by a virtual $\pi^0$, but is not important compared to the $\gamma\gamma$ channel due to phase-space suppression.}
The amplitude is calculated to be 
\begin{align}
{\cal M}(\phi\phi\to\gamma\gamma) & \,=\, \frac{\alpha_{\rm em}}{\pi} \frac{B_0^{}\, \big(C_{u\phi}^{S,uu}+C_{d\phi}^{S,dd}\big) }{m_{\pi^+}^2}  F(\rho_\pi)  \left[(q_1\cdot q_2) \big(\epsilon_1^*\cdot \epsilon_2^*\big)- \big(q_1\cdot \epsilon_2^*\big) \big(q_2\cdot \epsilon_1^*\big) \right]\,, ~~    
\end{align}
where $\alpha_{\rm em}$ denotes the fine structure constant, $\epsilon_{1,2}$ $(q_{1,2})$ are the photons' polarizations (momenta), and the loop function
\begin{align}
F(\rho_\pi) & \,=\, \frac{-1}{\rho_\pi} \left( \frac{1}{\rho_\pi}
\ln^2\frac{ \sqrt{\rho_\pi-4}-\sqrt{\rho_\pi} }{ \sqrt{\rho_\pi-4} +\sqrt{\rho_\pi} }
+1\right), &
\rho_\pi & \,=\, \frac{2q_1\cdot q_2}{m_{\pi^+}^2}
\,\approx\, \frac{4 m_\phi^2}{m_{\pi^+}^2} \,. & 
\end{align}
This translates into the annihilation rate 
\begin{align}
\langle\sigma v\rangle(\phi\phi\to \gamma\gamma)_{\tt I.D.}^{}
& \,\approx\, \frac{ B_0^2\, \big|C_{u\phi}^{S,uu}+C_{d\phi}^{S,dd}\big|^2 }{ 64\pi m_\phi^2 }
~ \frac{8\alpha_{\rm em}^2 m_\phi^4 }{\pi^2 m_{\pi^+}^4}|F(\rho_\pi)|^2 \,.
\end{align}
The amplitude for the photon mode is expected to receive a considerable correction from higher-order chiral terms. For a conservative estimation, we assume that this correction is, at most, of similar magnitude to the contribution from the pion loop to the cross section. In the numerical analysis, a factor of 2 is incorporated into the above equation to account for this effect. An ${\cal O}(1)$ factor difference will not substantially affect the limits shown in the plot.

The constraints on the pion and photon modes inferred in Ref.\,\cite{ODonnell:2024aaw} from the Planck CMB data turn out to exclude the light orange and light green areas depicted in the left panel of Fig.\,\ref{fig:dm}. 
As can be seen, the Planck CMB measurements impose very stringent bounds on the coupling  combination for \,$m_\phi \gtrsim 146$\,MeV.\, 
However, due to the loop suppression of the photon mode, the mass range \,$110 \lesssim m_\phi \lesssim 146$\,MeV\, remains viable for reproducing the correct DM relic density.  

\subsection{Collider searches}

Searches for DM have also been performed at colliders. 
Relevant to the effective interactions in Eq.\,(\ref{eq:LEFT}) are the recent  constraints acquired in Ref.\,\cite{Belyaev:2018pqr} from LHC monojet data. 
We have included the results in the left panel of Fig.\,\ref{fig:dm} (the dashed purple line) for comparison with the other limits.
Since the LHC monojet bound was obtained under the assumption that light quarks have flavor-universal couplings, the plotted region contains some minor uncertainties.

\subsection{Direct searches}

The DM-quark interactions given in \cref{eq:LEFT} can also generate signals in DM direct-detection experiments. 
For sub-GeV DM, the sensitivity of the conventional nuclear recoil searches rapidly diminishes because of detection threshold limitations, 
making it difficult to obtain meaningful constraints on the relevant parameter space. 
Nevertheless, it has been pointed out that the inelastic Migdal effect could offer a way to overcome this kinematic barrier and probe the low-mass region~\cite{Ibe:2017yqa}. 
This idea has been implemented in liquid xenon and argon experiments, including XENON1T~\cite{XENON:2019zpr}, DarkSide50~\cite{DarkSide:2022dhx}, LZ~\cite{LZ:2023poo}, and PandaX-4T~\cite{PandaX:2023xgl}, leading to significant bounds on the spin-independent DM-nucleon cross section. 

To apply these results to our DM scenario, we evaluate the corresponding cross section of DM-nucleon scattering in the zero-momentum-transfer limit.
For the interactions in Eq.\,(\ref{eq:LEFT}), only the flavor-diagonal operators contribute. 
By performing a nonrelativistic matching with the standard spin-independent DM-nucleon nonrelativistic operator \,${\cal O}_1^N \equiv 1_\phi 1_N$,\, where $N=\rm proton$ $(p)$ or neutron $(n)$, we arrive at the matching coefficients 
\begin{align}
c_1^N & \,=\, \frac{2 m_N^2}{m_u} f_{T_u}^{(N)} C_{u\phi}^{S,uu}  
+ \frac{2 m_N^2}{m_d} f_{T_d}^{(N)} C_{d\phi}^{S,dd} + \frac{2m_N^2}{m_s} f_{T_s}^{(N)} C_{d \phi}^{S,ss}
\nonumber \\
& \,=\, \frac{2m_N^2}{m_s} f_{T_s}^{(N)} \big( C_{u\phi}^{S,uu} + C_{d\phi}^{S,dd} \big) \big( r_+^N + r_-^N R_-^{} + R_s^{} \big) \,, & 
\end{align}
where $m_{u,d,s}$ denote the masses of the three light quarks ($u,d,s$) and $f_{T_{u,d,s}}^{(N)}$ represent the nucleon form factors associated with the scalar quark currents at zero-momentum transfer, which are defined by \,$f_{T_q}^{(N)} = (m_q/m_N)\langle N|\bar q q |N\rangle$.\,
In the second equality above, we have defined 
\begin{align}
r_\pm^N & \,=\, \frac{1}{2} \Bigg[ \frac{f_{T_u}^{(N)} m_s }{ f_{T_s}^{(N)} m_u } 
\pm \frac{f_{T_d}^{(N)} m_s }{ f_{T_s}^{(N)} m_d } \Bigg] \,, & 
R_-^{} & \,=\, \frac{C_{u \phi}^{S,uu}-C_{d \phi}^{S,dd} }{ C_{u\phi}^{S,uu}+C_{d\phi}^{S,dd}} \,, &   
R_s^{} & \,=\, \frac{C_{d \phi}^{S,ss} }{ C_{u\phi}^{S,uu}+C_{d\phi}^{S,dd}} \,.
\end{align}  
If the nucleon matrix elements respect isospin symmetry, \,$r_+^p=r_+^n\equiv r_+^{}$\, and \,$r_-^p = - r_-^n \equiv r_-^{}$.\,  
In the case of DM-quark couplings, a nonzero value of  $R_{-}$ implies that $c_1^p\neq c_1^n$, which corresponds to the isospin-violating DM scenario. 

By incorporating these isospin-violating couplings, the corresponding effective DM-nucleon cross section is calculated to be [see Eq.\,(H.37) in Ref.\,\cite{DelNobile:2021wmp}]
\begin{align}
\sigma_{\phi N} & \,=\, \frac{\mu_{\phi N}^2}{4\pi m_\phi^2} \bigg| \frac{m_N}{m_s} f_{T_s}^{(N)} \big(C_{u\phi}^{S,uu}+C_{d\phi}^{S,dd}\big) \bigg|^2
\big| r_+^{} + R_s^{} - (1 - 2Z/A ) r_-^{} R_-^{} \big|^2 \,, & 
\end{align}  
where $\mu_{\phi N}^{}$ is the reduced mass of the DM-nucleon system and $A$ ($Z$) is the atomic (mass) number of the target nucleus. 
Since the target material of xenon-based experiments such as PandaX-4T have several isotopes with abundances  summarized in Table\,1 of Ref.\,\cite{DelNobile:2021wmp}, we need to employ the weighted average \,$\sum_i \xi_i\,\sigma_{\phi N}^i$,\, where the sum runs over all isotopes $i$ with abundance $\xi_i$.  

Numerically, to establish our bound, we use the strictest limit on the effective cross section from PandaX-4T, represented by the red line for a heavy mediator in Fig.\,3 of Ref.\,\cite{PandaX:2023xgl}. 
For the nucleon form factors, we find that $r_\pm$ are connected to the nucleon isovector charge $g_S^{u-d}$ and the nucleon $\sigma$-terms, $\sigma_{\pi N}$ and $\sigma_s$ \cite{FlavourLatticeAveragingGroupFLAG:2024oxs}, through the relations   
\,$r_+ = (m_s/2\sigma_s)(\sigma_{\pi N}/m_{ud})$\, and \,$r_-^{} = (m_s/2\sigma_s) g_S^{u-d}$,\, where \,$m_{ud}\equiv (m_u+m_d)/2$.\, 
Furthermore, \,$(m_N/m_s)f_{T_s}^{(N)} = \sigma_s/m_s$.\, 
The parameters $g_S^{u-d}$, $\sigma_{\pi N}$, and $\sigma_s$ currently have large uncertainties in lattice computations. 
For our illustration, we adopt the latest 
FLAG~\cite{FlavourLatticeAveragingGroupFLAG:2024oxs} global average values with three flavors:
$g_S^{u-s}=1.083(69)$,\, 
$\sigma_{\pi N}=42.2(2.4)$\,MeV,  and \,$\sigma_s=44.9(6.4)$\,MeV\, from Eqs.\,\,(428), (448), and (450), respectively, in Ref.\,\cite{FlavourLatticeAveragingGroupFLAG:2024oxs}.
For the quark masses and their ratios, we employ \,$m_s/m_{ud}=27.33_{-0.14}^{+0.18}$\,
and \,$m_s = 93.5(8)$\,MeV\, from Ref.\,\cite{ParticleDataGroup:2024cfk}. 
The above numbers lead to \,$\sigma_s/m_s=0.48(7)$,\, $r_+=12.8(2.0)$,\, 
and \,$r_-=1.13(18)$.\,
In the right panel of \cref{fig:dm}, we 
present the permitted regions on the $R_-$-$R_s$ plane for three benchmark DM masses, $m_\phi=(120,130,140)$\,MeV, that fulfill the restrictions inferred from DM indirect and collider searches. 
As displayed in this figure, to comply with the Migdal-effect limits, 
$R_-$ must significantly deviate from unity,
indicating that substantial isospin violation is needed.
It is worth noting that the Migdal-effect constraints are affected by uncertainties in the form factors, which may cause slight shifts in the allowed parameter space.

\section{Conclusions\label{concl}}

We have explored the possibility that the recently measured excess of \,$B^+\to K^+$+$\slashed E$\, rate relative to its SM expectation originates from DM-quark interactions that can also saturate the potential new-physics window in the newly observed \,$K^+\to\pi^+$+$\slashed E$.\,  
Starting with effective operators involving quarks and the real scalar boson $\phi$ acting as the DM, we investigate the ranges of their couplings that can satisfy the preceding requirements, taking into account the constraints from the DM sector. 
In particular, restrictions from the relic density data and from DM direct, indirect, and collider searches turn out to restrain the $\phi$ mass to be between 110 and~146~MeV. 
We further find that the NA62 data on \,$K^+\to\pi^+$+$\slashed E$,\, including the information on its signal regions, reduce this range to 
\,$110\;\mbox{\small$\lesssim$}\; m_\phi \;\mbox{\small$\lesssim$}\; 130$~MeV.\,
On the upside, this implies that the $\phi$ interactions for \,$130 \;\mbox{\small$\lesssim$}\; m_\phi \;\mbox{\small$\lesssim$}\; 146$~MeV\, remain unconstrained.
Improved data from future measurements of \,$B\to K$+$\slashed E$\, by Belle II, \,$K^+\to\pi^+$+$\slashed E$\, by NA62, and \,$K_L\to\pi^0$+$\slashed E$\, by KOTO and from various upcoming DM searches can be expected to test our new-physics scenario more stringently.
It goes without saying that more precise predictions for these decay modes in the SM and better estimation of the Migdal effect are desirable for a sharper comparison of experiment and theory.  

\section*{Acknowledgments}

J.T. and G.V. thank the Tsung-Dao Lee Institute, Shanghai Jiao Tong University, for kind hospitality and support during the preparation of this paper. 
X.-G.H. was supported by the Fundamental Research Funds for the Central Universities, by the National Natural Science Foundation of the People’s Republic of China (No.\,12090064, No.\,11735010, No.\,12205063, No.\,11985149, No.\,12375088, and No.\,W2441004), and by MOST\,109–2112-M-002–017-MY3. 
X.-D.M. was supported by Grant No.\,NSFC-12305110.

\newpage
\setlength{\bibsep}{.15\baselineskip plus 0.1ex}
 
\bibliographystyle{utphys.bst} 
\bibliography{refs.bib}

\end{document}